\documentclass[conference]{IEEEtran}
\IEEEoverridecommandlockouts

\usepackage{cite}
\usepackage{graphics}
\usepackage{graphicx}
\graphicspath{{Graphs/}}

\usepackage[cmex10]{amsmath}
\usepackage{amssymb}
\usepackage{algorithmic}
\usepackage{array}
\usepackage{mdwmath}
\usepackage{mdwtab}
\usepackage[tight,footnotesize]{subfigure}

\begin{document}

\title{Robust Fitting of Ellipses and Spheroids\thanks{This research was supported in part by the U.S. Office of Naval
Research under grant numbers N00014-07-1-0555 and N00014-09-1-0342, and in part by the U.S. Army Research Office under grant number  
W911NF-07-1-0185.}}

\author{\IEEEauthorblockN{Jieqi Yu}
\IEEEauthorblockA{Department of Electrical Engineering\\Princeton University\\
Princeton, NJ, 08544\\
Email: jieqiyu@princeton.edu\\
Phone: 609-258-6868}
\and \IEEEauthorblockN{Sanjeev R. Kulkarni}
\IEEEauthorblockA{Department of Electrical Engineering\\Princeton University\\
Princeton, NJ, 08544\\
Email: kulkarni@princeton.edu}
\and
\IEEEauthorblockN{H. Vincent Poor}
\IEEEauthorblockA{Department of Electrical Engineering\\Princeton University\\
Princeton, NJ, 08544\\
Email: poor@princeton.edu}}

\maketitle

\begin{abstract}
Ellipse and ellipsoid fitting has been extensively researched and widely applied. Although traditional fitting methods provide accurate estimation of ellipse parameters in the low-noise case, their performance is compromised when the noise level or the ellipse eccentricity are high. A series of robust fitting algorithms are proposed that perform well in high-noise, high-eccentricity ellipse/spheroid (a special class of ellipsoid) cases. The new algorithms are based on the geometric definition of an ellipse/spheroid, and improved using global statistical properties of the data. The efficacy of the new algorithms is demonstrated through simulations.

\end{abstract}

\section{Introduction}
Ellipse and ellipsoid fitting has been extensively studied and has broad applications. For example, an ellipse can serve as a geometric primitive model in computer vision and pattern recognition, which allows reduction and simplification of data for higher level processing. Ellipse fitting is also essential in applied sciences from observational astronomy (orbit estimation) to structural geology (strain in natural rocks). Moreover, ellipsoid fitting (e.g., minimum volume ellipsoid estimator \cite{Rousseeuw1987}) is a useful tool in outlier detection.

In this paper, we consider the following issues: In Section \ref{background}, the ellipse fitting problem is formulated and several important algorithms are reviewed. In Section \ref{Obj}, a new objective function based on the geometric definition of ellipse is introduced, and it is further extended to three ellipse fitting algorithms in Section \ref{algorithms}. Then a spheroid fitting algorithm is proposed in Section \ref{SecSpheroid}, which is followed by the experimental results in Section \ref{SimuResult}.

\section{Formulation and Background}\label{background}
As a classical problem, ellipse fitting has a rich literature. Various algorithms have been proposed from very different perspectives. We start our discussion by reviewing several classes of the most important ellipse fitting algorithms.

The problem of ellipse fitting can be modeled as follows. We have data points $\{\mathbf{z}_i=(x_i,y_i)\}_{i=1}^n$, which are points on an ellipse corrupted by noise. The objective is to fit an ellipse with unknown parameters $\mathbf{\beta}$ to the data points, so that the total error is minimized. The measure of error differs for different classes of algorithms.

The most intuitive class of ellipse fitting algorithms is \emph{algebraic fitting}, which uses \emph{algebraic error} as a measure of error. An ellipse can be described by an implicit function $P(\mathbf{\beta})=Ax^2+Bxy+Cy^2+Dx+Ey+F=0$, where $\mathbf{\beta}=(A,B,C,D,E,F)$ denotes the ellipse parameters. The algebraic error from a data point $\mathbf{z}_i=(x_i,y_i)$ to the ellipse is thus $Ax_i^2+Bx_iy_i+Cy_i^2+Dx_i+Ey_i+F$. 

The most efficient and widely used algorithm in this category was proposed by Fitzgibbon et al. \cite{Fitzgibbon1999}. It is a least squares optimization problem with respect to the algebraic criterion:
\begin{eqnarray}
           \min_\mathbf{\beta} && \sum^n_{i=1}(Ax_i^2+Bx_iy_i+Cy_i^2+Dx_i+Ey_i+F)^2 \\
 \nonumber \text{s.t.} && B^2-4AC=1.
\end{eqnarray}
The constraint ensures that the optimization result is an ellipse instead of a general conic section and also prevents the problem of parameter free scaling. This optimization problem is further reduced to a generalized eigenvalue problem, which can be efficiently solved. In addition, Halir et al. improved the algorithm into a numerically stable version \cite{Halir1998}. Algebraic fitting has the advantage of low computational complexity. However, the algebraic criterion lacks geometric interpretation, and the algorithm is difficult to generalize to three dimensions due to its non-linear constraint.

To overcome the shortcomings of algebraic fitting, Ahn et al. proposed \emph{orthogonal least squares fitting} (OLSF) \cite{Ahn2001}. OLSF minimizes the sum of squared \emph{Euclidian}
distance, which is defined as the orthogonal distance from the data point to the ellipse:
\begin{equation}
\min_\mathbf{\beta} \sum^n_{i=1}\|\mathbf{z}_i-\mathbf{z}_i'(\mathbf{\beta})\|^2,
\end{equation}
where $\mathbf{z}_i'(\mathbf{\beta})$ is the orthogonal contacting point, which is the point on the ellipse that has the shortest distance to the corresponding data point. OLSF has a clear geometric interpretation and features high accuracy. Moreover, it can be generalized to the three-dimensional case \cite{Ahn2002}. Unfortunately, OLSF is computationally intensive. It employs the iterative Gauss-Newton algorithm, and in each iteration, the orthogonal contacting points have to be found for each data point, being iterative itself.

Various extensions to OLSF algorithms have also been proposed. Angular information is incorporated into the OLSF algorithm in Watson's 2002 paper \cite{WatsonDec.2002}, in which the orthogonal geometric distance is replaced by the distance along the known measurement angle. Moreover, instead of the $l_2$ norm in OLSF, $l_1$, $l_\infty$ and $l_p$ norms have been considered as well \cite{Watson2002} \cite{Al-Subaihi2005} \cite{Atieg2003}.

The third class of algorithms consists of \emph{Maximum Likelihood} (ML) algorithms, which were proposed in \cite{Chojnacki2000} and \cite{LeedanJun.2000}. The key steps of the two ML algorithms, the fundamental numerical scheme \cite{Chojnacki2000} and the heteroscedastic errors-in-variables scheme \cite{LeedanJun.2000}, have been proven to be
equivalent in \cite{Chojnacki2004}.

In \cite{Chojnacki2000}, Chojnacki et al. assume that the data points are independent and have a multivariate normal distribution: $\mathbf{z}_i\sim \mathcal{N}(\mathbf{z}'_i,\Lambda_{\mathbf{z}_i})$. The ML solution is then reduced to an optimization problem based on \emph{Mahalanobis distance}:
\begin{equation}
\min_{\mathbf{\beta}}\sum^n_{i=1}(\mathbf{z}_i-\mathbf{z}'_i(\mathbf{\beta}))^T\Lambda^{-1}_{\mathbf{z}_i}(\mathbf{z}_i-\mathbf{z}'_i(\mathbf{\beta})).
\end{equation}
The fundamental numerical scheme is implemented to solve a variational equation iteratively by solving an eigenvalue problem at each iteration until it converges. The ML algorithms are accurate with moderate computational cost. However, when the noise is large or the eccentricity of the ellipse is large, the algorithm breaks down, because when any data point is close to the center of the estimated ellipse, one of the matrices in the algorithm has elements that tend to infinity.

All three classes of algorithms described above have their advantages and perform well when the noise level is low. However, they share a common disadvantage of not being robust enough for highly noisy data. In this paper, we propose a series of algorithms that are resistant to large noise, and can be generalized to three-dimensions easily, with competitive accuracy and moderate computational cost.

\section{A New Geometric Objective Function}\label{Obj}
Note that both OLSF and ML algorithms estimate the nuisance parameters $\mathbf{z}_i'$ (point on the ellipse that generates the data point) in addition to the parameters of the ellipse. It is desirable to bypass these nuisance parameters and have a more intrinsic fitting method with a clear geometric interpretation.

Recall the geometric definition of an ellipse. An ellipse is the locus of points such that the sum of the distances from that point to two other fixed points (called the foci of the ellipse) is constant. I.e., $\mathbf{z}$ is a point on the ellipse if and only if
\begin{equation}
\|\mathbf{z}-\mathbf{c_1}\|+\|\mathbf{z}-\mathbf{c_2}\|=2a,
\end{equation} where $\|\cdot\|$ denotes the $l_2$ norm, $\mathbf{c_1}$ and $\mathbf{c_2}$ are the two foci, and $a$ is the length of the semi-major axis.

Based on the geometric definition, the ellipse fitting problem can be naturally formulated as an optimization problem with a new geometric objective function:
\begin{equation}\label{NewObjective}
    \min_{\mathbf{c}_{1},\mathbf{c}_{2},a}\frac{1}{n}\sum_{i=1}^{n}(\|\mathbf{z}_{i}-\mathbf{c}_{1}\|+\|\mathbf{z}_{i}-\mathbf{c}_{2}\|-2a)^{2},
\end{equation} where $n$ denotes the number of data points.

The objective function in (\ref{NewObjective}) has several advantages. First, it has a clear geometric interpretation: it is the expected squared difference between the sum of the distances from the data points to the foci and the length of the major axis. Second, the parameters in the objective function, $\mathbf{c}_1$, $\mathbf{c}_2$ and $a$, are intrinsic parameters of ellipses, which are translation and rotation invariant. The third advantage is that the objective function is ellipse specific, and thus no extra constraints are needed. As a result, the minimization problem can be readily solved by gradient descent algorithms. Lastly, and most importantly, the objective function possess one more property that contributes to the robustness of our algorithm, which is demonstrated by comparing with the objective function of OLSF.

\begin{figure}[!t]
\centering
\includegraphics[width=1.6in]{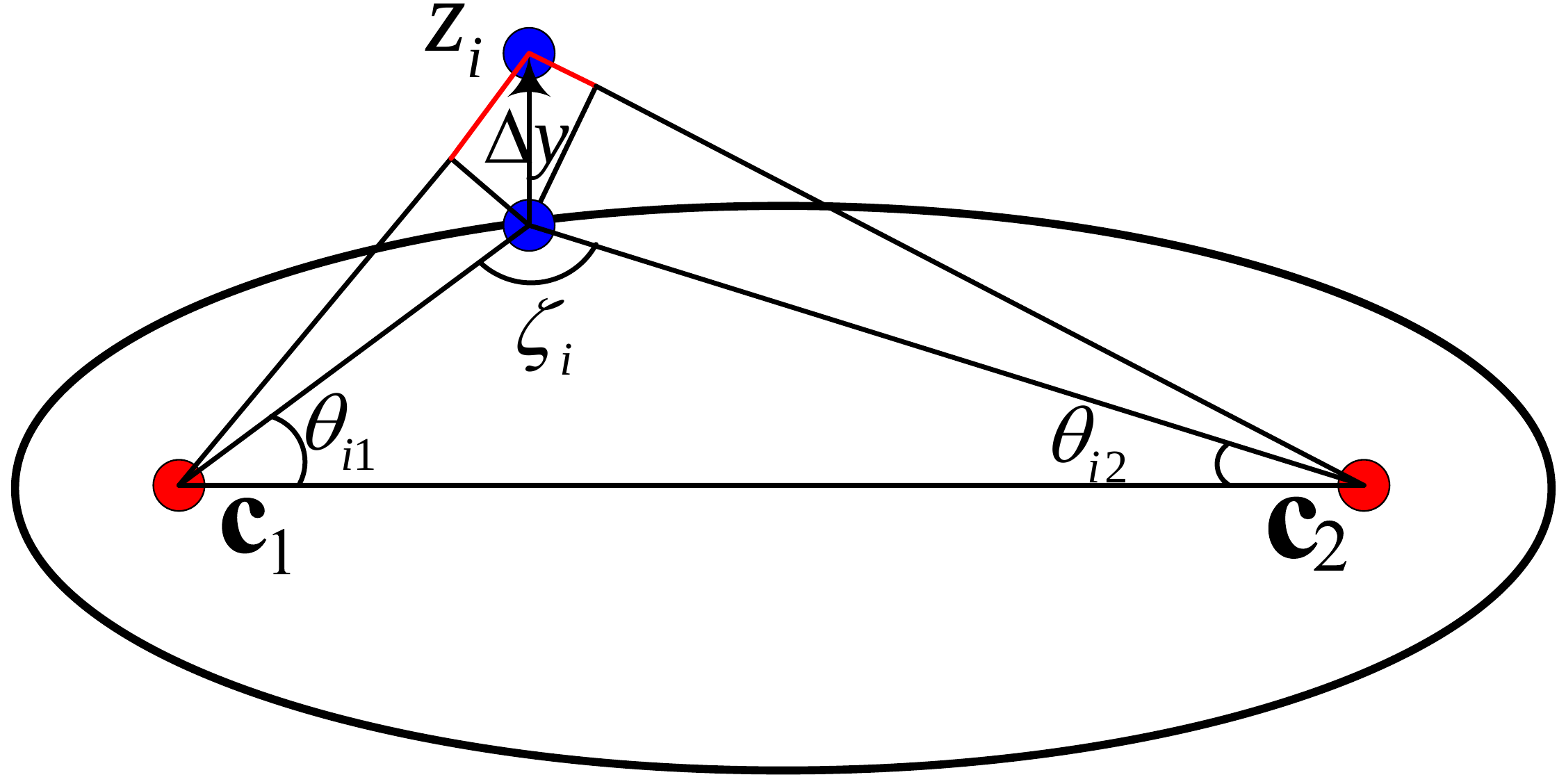}~~
\includegraphics[width=1.6in]{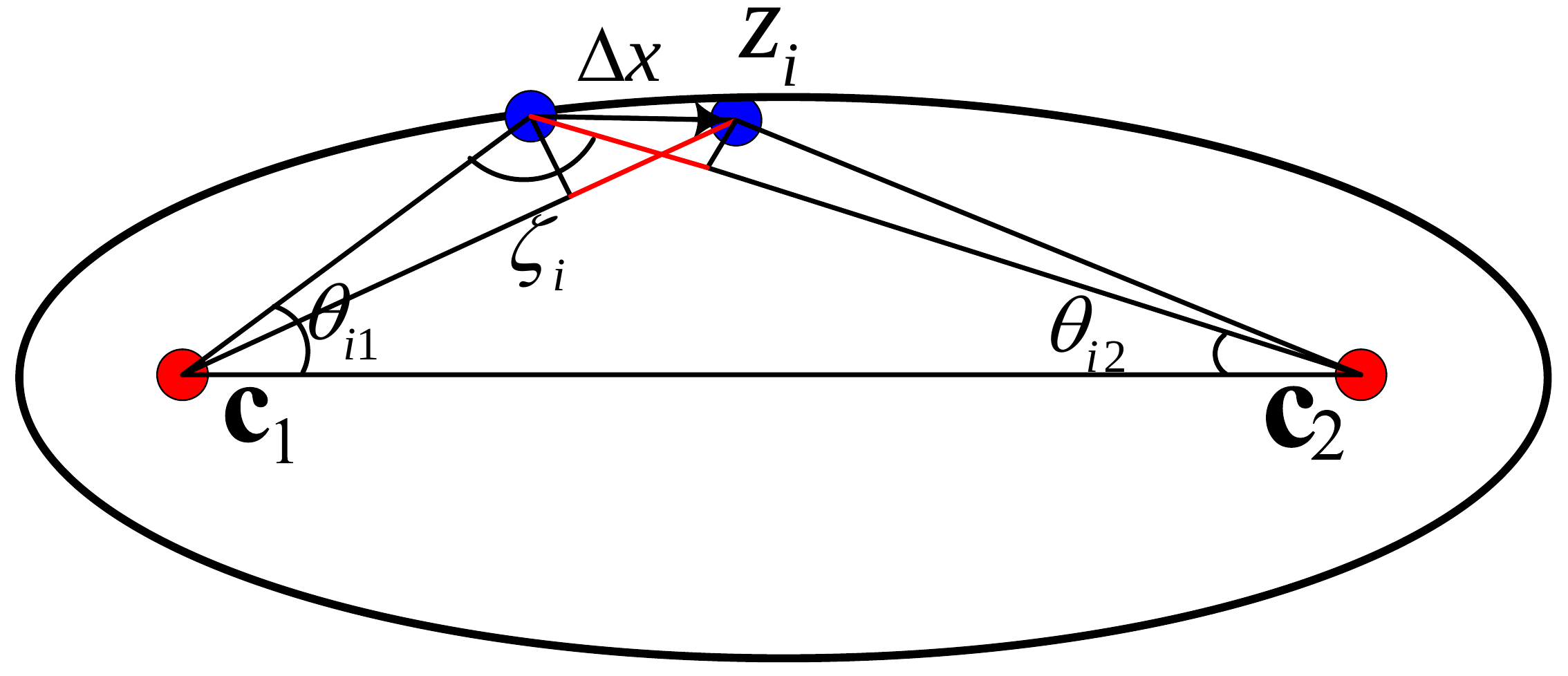}
\caption{Derivation of the expected contribution of a noisy data point to our objective function.} \label{analysisgraph}
\end{figure}

Assume that the data points are independent and have a multivariate normal distribution: $\mathbf{z}_i\sim \mathcal{N}(\mathbf{z}'_i,\Lambda_{\mathbf{z}_i})$. The expected contribution of a noisy data point to the objective function of OLSF is approximately
\begin{equation} 
\mathbf{E}(||\mathbf{z}_i-\mathbf{z}'_i||^2) \approx \sigma^2. 
\end{equation}
Note that this is homogeneous for all the data points. However, for our objective function, the expected contribution can be approximated as (see Fig. \ref{analysisgraph})
\begin{eqnarray}\label{analysis}
	\mathbf{E}[f_{\mathbf{z}_i}] &=& \mathbf{E}\left[\left(\Delta y(\sin \theta_{i1}+\sin\theta_{i2})+\Delta x(\cos_{\theta_{i1}}-\cos_{\theta_{i2}})\right)^2\right]\nonumber\\
	&=& 2\sigma^2 + 2\sigma^2(\sin\theta_{i1}\sin\theta_{i2}-\cos\theta_{i1}\cos\theta_{i2})\nonumber\\
	&=& 2\sigma^2(1-\cos(\theta_{i1}+\theta_{i2}))\nonumber\\
	&=& 2\sigma^2(1+\cos\zeta_i),
\end{eqnarray}
where $\zeta_i$ is the angle $\angle \mathbf{c}_1 \mathbf{z}_i \mathbf{c}_2$. This quantity is heterogeneous around the periphery of the ellipse: the objective function puts a large weight on the data points located at two ends of the major axis. As a result, our algorithm provides a highly robust and accurate major \emph{axis orientation} estimate.

However, the objective function in (\ref{NewObjective}) has a global minimum at infinity. When the foci move away from each other toward infinity and the semi-major axis length tends to infinity, the value of the objective function approaches zero. So when the noise level is high, the local minimum (that we desire in this case) is smeared out by the noise, resulting in the estimated foci slipping away along the major axis.

In order to take advantage of the robustness of ellipse axis orientation estimation while overcoming the problem of having a global minimum at infinity, we propose three modified algorithms in the next section, so as to increase the robustness and accuracy of the algorithm directly using our new objective function.

\newpage
\section{Three Modified Algorithms for \\Ellipse Fitting}\label{algorithms}
In this section, we introduce three modified algorithms based on the objective function proposed in Section \ref{Obj}. Each of the three algorithms overcomes the problem of a global minimum at infinity and achieves a robust and accurate result.

\subsection{Penalized Objective Function}
The most intuitive way to eliminate the global minimum at infinity is penalization. The penalty term should favor smaller semi-major lengths, so that the foci will not diverge. As a result, we propose the following penalized objective function:
\begin{equation}\label{Penalized}
 \frac{1}{n}\sum_{i=1}^{n}(||\mathbf{z_{i}}-\mathbf{c_{1}}||+||\mathbf{z_{i}}-\mathbf{c_{2}}||-2a)^{2}+\lambda \hat{a}_{\max}\hat{\sigma}\exp\left(\left(\frac{a}{\hat{a}_{\max}}\right)^{4}\right),
\end{equation}
where the second term is the penalty term, $\lambda$ is a tuning parameter, $\hat{a}_{\max}$ is an estimated upper bound for the semi-major length $a$, and $\hat{\sigma}$ is an estimate of the noise standard deviation. $\hat{\sigma}$ and $\hat{a}_{\max}$ are estimated during the initialization procedure.

The term $\exp((\frac{a}{\hat{a}_{\max}})^{4})$ tends to infinity rapidly when $a$ exceeds $\hat{a}_{\max}$, thus eliminating the global minimum at infinity. The penalty term is also proportional to the estimated noise level. When the noise level is high, the penalty term is large, so as to increase the robustness of the algorithm. On the other hand, when the noise level is low, the penalty term is small to ensure a small bias added to the objective function. The coefficient $\hat{a}_{\max}$ is a scaling factor to accommodate the size of the ellipse we are fitting.

In the initialization, $\hat{a}_{\max}$ is estimated as the largest distance from a data point to the mean of all the data points $\mathbf{z}_{\text{mean}}$. $a$ is initialized as the mean of the distance from the data points to $\mathbf{z}_{\text{mean}}$. To estimate $\hat{\sigma}$, we run the gradient descent algorithm for (\ref{NewObjective}) and terminate when $a$ exceeds $\hat{a}_{\max}$. Then $\hat{\sigma}$ is estimated as the square root of the resulting objective function value. This simple noise level estimation method suffices for our purposes here.

The penalized optimization problem (\ref{Penalized}) can be readily solved by gradient descent algorithms once initialized. The global minimum at infinity is eliminated and the resulting algorithm has good accuracy and high robustness, which we will show in Section \ref{SimuResult} via simulation results.

\subsection{Axial Guided Ellipse Fitting}
The second algorithm we propose to overcome the problem described in Section \ref{Obj} is \emph{axial guided fitting}. Recall that the major axis orientation estimate is accurate and robust. So the only thing left to be determined is the size of the ellipse, i.e., the semi-major length $a$ and semi-minor length $b$.

To estimate $a$ and $b$, we first solve (\ref{NewObjective}) to estimate the major-axis orientation $\phi$ and ellipse center $(x_{c},y_{c})$. Then we translate and rotate the data points $\{(x_{i},y_{i})\}_{i=1}^n$ so that the estimated ellipse is in the standard position (centered at the origin with its major axis coinciding with the x-axis).  The resulting data points are
\begin{equation}
        x_{i}' = \cos\phi(x_{i}-x_{c})+\sin\phi(y_{i}-y_{c})
\end{equation}
and
\begin{equation}
        y_{i}' = -\sin\phi(x_{i}-x_{c})+\cos\phi(y_{i}-y_{c}).
\end{equation}

Then the semi-major length $a$ is determined such that a certain percentage, $P_{a}$, of the data points satisfy $x_i'\in[x_{c}-a,x_{c}+a]$. The semi-minor length $b$ is determined in a similar manner with percentage $P_b$. Naturally, $P_{a}$ and $P_{b}$ are related to the noise level. If we make additional assumptions that the data points are independent of each other and have a multivariate normal distribution: $\mathbf{z}_i\sim \mathcal{N}(\mathbf{z}'_i,\Lambda_{\mathbf{z}_i})$, and that they are distributed uniformly in angle around the periphery, then the relationship between $P_{a}$, $P_{b}$ and the noise level $\sigma$ can be approximated as
\begin{equation}
P_a(\gamma)\approx\frac{1}{\pi}\int_{0}^{\frac{\pi}{2}}\left(\text{erf}\left(\frac{1}{\sqrt{2}}\gamma(1-\cos\theta)\right)\right) \mathrm{d}\theta,\label{Pa}
\end{equation}
and
\begin{equation}
P_b(\gamma')\approx\frac{1}{\pi}\int_{0}^{\frac{\pi}{2}}\left(\text{erf}\left(\frac{1}{\sqrt{2}}\gamma'(1-\sin\theta)\right)\right) \mathrm{d}\theta,\label{Pb}
\end{equation}
where $\gamma=\frac{a}{\sigma}$ and $\gamma'=\frac{b}{\sigma}$. Recall that we have the approximation (\ref{analysis}). Assuming that the estimated ellipse is close to the true model, the noise level can be readily estimated as
\begin{equation}\label{NoiseEst}
\sigma^2 = \frac{1}{n} \sum_{i=1}^{n} \frac{f_{\mathbf{z}_i}}{ 2(1+\cos\zeta_i)}.
\end{equation}

With this noise estimate, we can perform the axial guided fitting as described, which results in the following procedure:
\begin{enumerate}
\item Solve (\ref{NewObjective}) by gradient descent algorithms to obtain $\phi$, $(x_c,y_c)$, $f_{\mathbf{z}_i}$ and $\zeta_i$, $\forall i$;
\item Translate and rotate the data set so that the estimated ellipse is located at the standard position;
\item Estimate the noise level by (\ref{NoiseEst});
\item Calculate $P_a$ and $P_b$ from (\ref{Pa}) and (\ref{Pb});
\item Find $a$ and $b$ so that a portion $P_a$ of the data points satisfy $x_i'\in[x_c-a,x_c+a]$ and a portion $P_b$ of the data points satisfy $y_i'\in[y_c-b,y_c+b]$.
\end{enumerate}

Axial guided ellipse fitting divides the ellipse fitting problem into two stages: orientation estimation and size estimation. In applications where the noise level is known a priori, axial guided fitting could be simplified and becomes more efficient and accurate.

\subsection{Weighted Objective Function}
In order to take advantage of the robust ellipse orientation estimation and obtain an accurate size estimation as well, we propose the following \emph{weighted objective function}:
\begin{equation}\label{weighted}
\min_{\mathbf{c}_{1},\mathbf{c}_{2},a}\frac{1}{n}\sum_{i=1}^{n}\frac{1}{1+\beta\cos\zeta_i}(||\mathbf{x}_{i}-\mathbf{c}_{1}||+||\mathbf{x}_{i}-\mathbf{c}_{2}||-2a)^{2},
\end{equation}
where $\beta$ is a tuning parameter which varies from 0 to 1. 

When $\beta=0$, (\ref{weighted}) is the same as the original objective function, so that we can obtain accurate ellipse orientation estimation. On the other hand, when $\beta=1$, the angle dependent weight $1+\cos\zeta_i$ is applied to mimic the objective function of OLSF so as to obtain an accurate size estimation. By varying $\beta$ from $0$ to $1$ gradually, we avoid stray local minima at first, and aim for accuracy in the end. 

The weighted objective minimization problem is solved by gradient descent in such a way that $1+\beta\cos\zeta_i$, $\forall i$ is assumed to be constant in each iteration and updated afterwards. The parameter $\beta$ is updated in a linear manner. We will show the efficacy of this algorithm in Section \ref{SimuResult}.

\section{Spheroid Fitting}\label{SecSpheroid}
So far, we have proposed three modified algorithms based on the geometric definition of an ellipse. In this section, we generalize our method to the three-dimensional case.

Unfortunately, general ellipsoids do not have a natural geometric definition similar to ellipses. Nonetheless, we can still generalize our algorithm to three-dimensions in the case of a spheroid. A spheroid is defined as a quadric surface obtained by rotating an ellipse about one of its principal axes; in other words, a spheroid is an ellipsoid with two equal semi-diameters. Although a spheroid is a special case of an ellipsoid, this special case may have broad applications.

According to the definition, a spheroid has the same basic geometric property as an ellipse. This means that all algorithms proposed in Section \ref{Obj} and \ref{algorithms} can be easily generalized to three dimensions in the natural way. For example, in the case of the weighted objective function, we still have the optimization problem
\begin{equation}
\min_{\mathbf{c}_{1},\mathbf{c}_{2},a}\frac{1}{n}\sum_{i=1}^{n}\frac{1}{1+\beta\cos\zeta_i}(||\mathbf{x}_{i}-\mathbf{c}_{1}||+||\mathbf{x}_{i}-\mathbf{c}_{2}||-2a)^{2},\label{spheroid}
\end{equation}
with $\mathbf{z}_i$, $\mathbf{c}_1$, $\mathbf{c}_2 \in \mathbb{R}^3$.

We now have a group of algorithms almost the same as in the ellipse fitting case, and we will demonstrate the fitting result of (\ref{spheroid}) as an example at the end of the next section.

\section{Experimental Results}\label{SimuResult}
To demonstrate the efficacy of the algorithms described above, we describe a series of experiments in different settings. Synthetic data has been used for the simulations.

A set of points on the perimeter of the ellipse were drawn according to a uniform distribution in angle. The data points were obtained by corrupting the true points with independent and identically distributed (i.i.d.) additive Gaussian noise, with mean $\mathbf{0}$ and covariance $\sigma^2\mathbf{I}$. The error rate is defined as the normalized area difference between the true ellipse $E_t$ and the fitted ellipse $E_f$:
\begin{equation}
\text{error rate}=\frac{S_{E_{t}\bigcup E_{f}}-S_{E_{t}\bigcap E_{f}}}{2S_{E_{t}}},
\end{equation}
where $S_{E_{f}\bigcup E_{f}}-S_{E_{f}\bigcap E_{f}}$ is the area difference and $S_{E_{t}}$ denotes the area of the true ellipse.

\subsection{Comparison of the proposed algorithms}
The accuracy of the method based on the original objective function, penalized fitting, axial guided fitting and weighted objective fitting were tested and compared under a wide range of noise levels.

\begin{figure}[!t]
\centering
\includegraphics[width=2.6in]{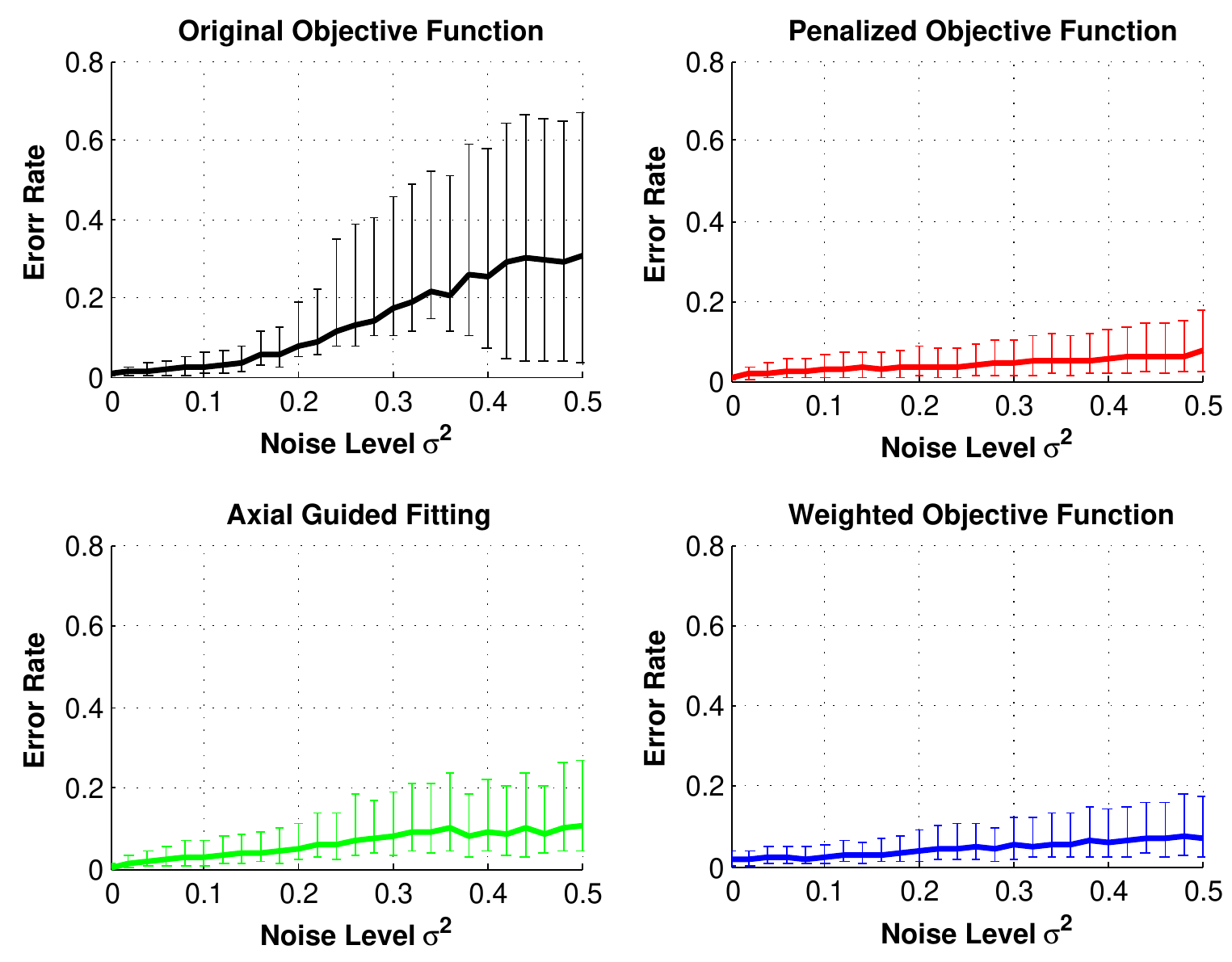}
\caption{Accuracy of the algorithms: average error rate under a wide range of noise levels for the four proposed algorithms. The lower bound and upper bound of the error bars are the 20\% and 80\% quantiles of 50 trials.} \label{Comparison}
\end{figure}

We consider an ellipse in standard position (centered at the origin and without rotation) with semi-major and semi-minor lengths $5$ and $3$ respectively. Fifty data points were drawn for each trial for a total run of $50$ trials per noise level. Fig. \ref{Comparison} shows the mean error rate for each algorithm under a range of noise levels ($\sigma^2$ from $0$ to $0.5$). The lower bounds of the error bars are the 20 percent quantiles and the upper bounds are the 80 percent quantiles of the $50$ trials.

As shown by the simulation results, the three revised algorithms exhibit substantial improvement compared with the method using the original objective function. Penalized fitting and weighted objective fitting have comparable performance, which is slightly better than that of axial guided fitting.

\begin{figure}[!t]
\centering
\includegraphics[width=1.9in]{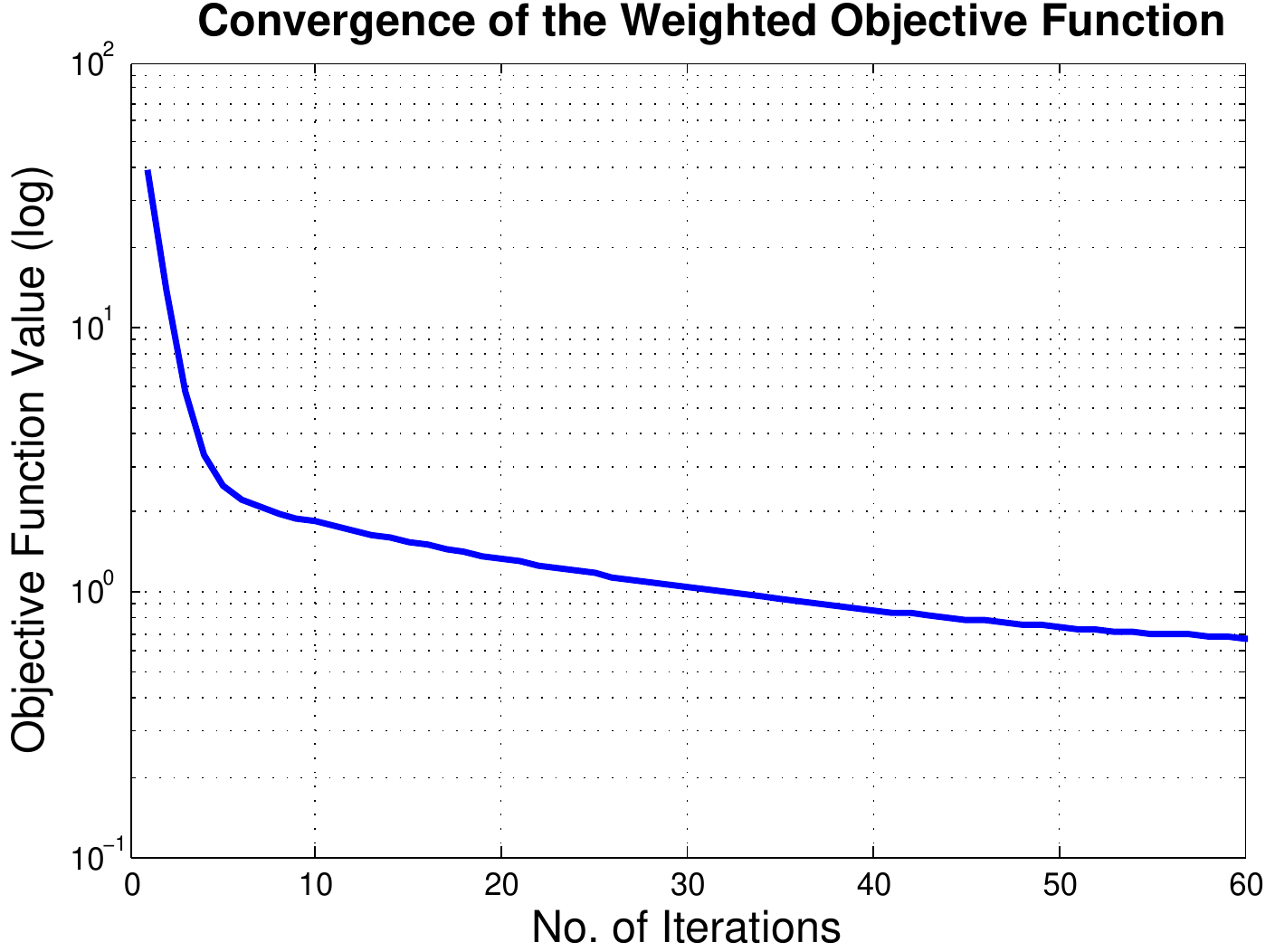}
\caption{Convergence rate of weighted objective fitting (log scale).} \label{Convergence}
\end{figure}

Fig. \ref{Convergence} shows the convergence rate of weighted objective fitting. Note that the plot is on a log scale. So the algorithm converges faster than exponential in the first few iterations. As for penalized fitting and axial guided fitting, they converge at similar speeds, except that penalized fitting has an initialization procedure and axial guided fitting needs noise estimation.

\subsection{Comparison with Algebraic Fitting, OLSF, and ML}
In this subsection, we describe simulations to compare our method with the previous methods (algebraic fitting, OLSF and ML), in terms of accuracy, under a high-eccentricity situation over a wide range of noise levels. Penalized fitting is used as a representative for the proposed algorithms. For algebraic fitting, OLSF and ML algorithms, we implemented the numerically stable version of algebraic fitting from \cite{Halir1998}, the OLSF algorithm proposed by Ahn et al. from \cite{Ahn2001}, and the FNS ellipse fitting algorithm from \cite{Chojnacki2000} respectively.

An ellipse in standard position with semi-major and semi-minor lengths $8$ and $2$, respectively, were used in the simulations. Fifty data points were drawn for each trial for a total run of $50$ trials per noise level for penalized fitting, algebraic fitting and OLSF, and a total run of $1000$ trials for FNS.

\begin{figure}[!t]
\centering
\includegraphics[width=2.4in]{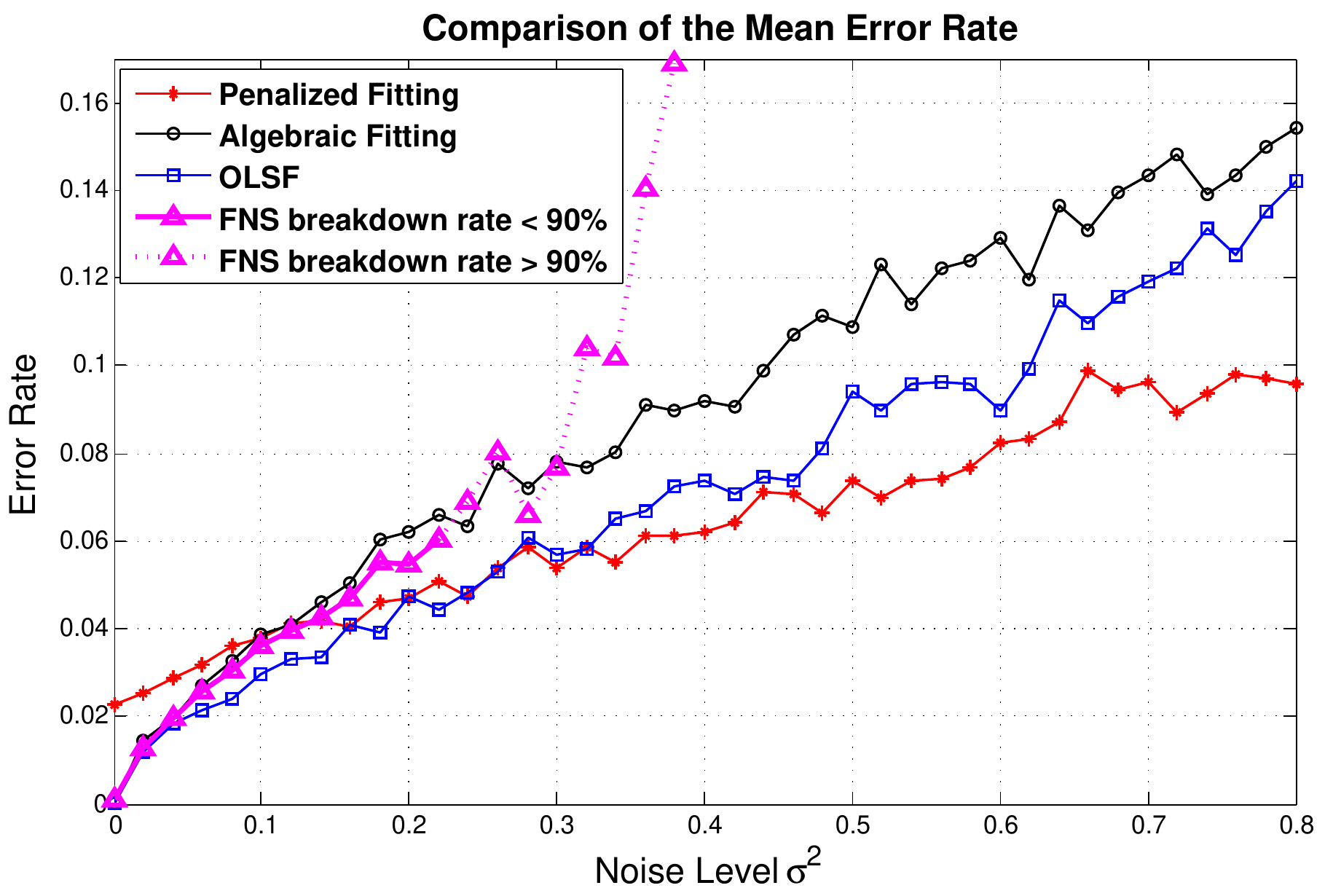}
\caption{Average error rate under a wide range of noise levels for algebraic fitting, OLSF, FNS and penalized fitting} \label{ComparisonMean}
\end{figure}

Fig. \ref{ComparisonMean} shows the mean error rate under a range of noise levels ($\sigma^2$ from $0$ to $0.8$) for the four algorithms. Although penalized fitting performs slightly worse than the other three algorithms when the noise level is low, it outperforms them when the noise level increases, which shows the robustness of our algorithms. The curve with triangle markers represents the FNS algorithm. It has high accuracy when the noise level is low, yet it breaks down when there are data points near the center of the estimated ellipse, which happens often for moderate or high noise levels. The FNS curve represrents the average error on those trials (out of 1000 trials) for which the algorithm produced an estimate.  The dotted segment of the FNS plot indicates that the algorithm failed to produce an estimate for more than $90\%$ of the trials.

\begin{figure}[!t]
\centering
\includegraphics[width=2.3in]{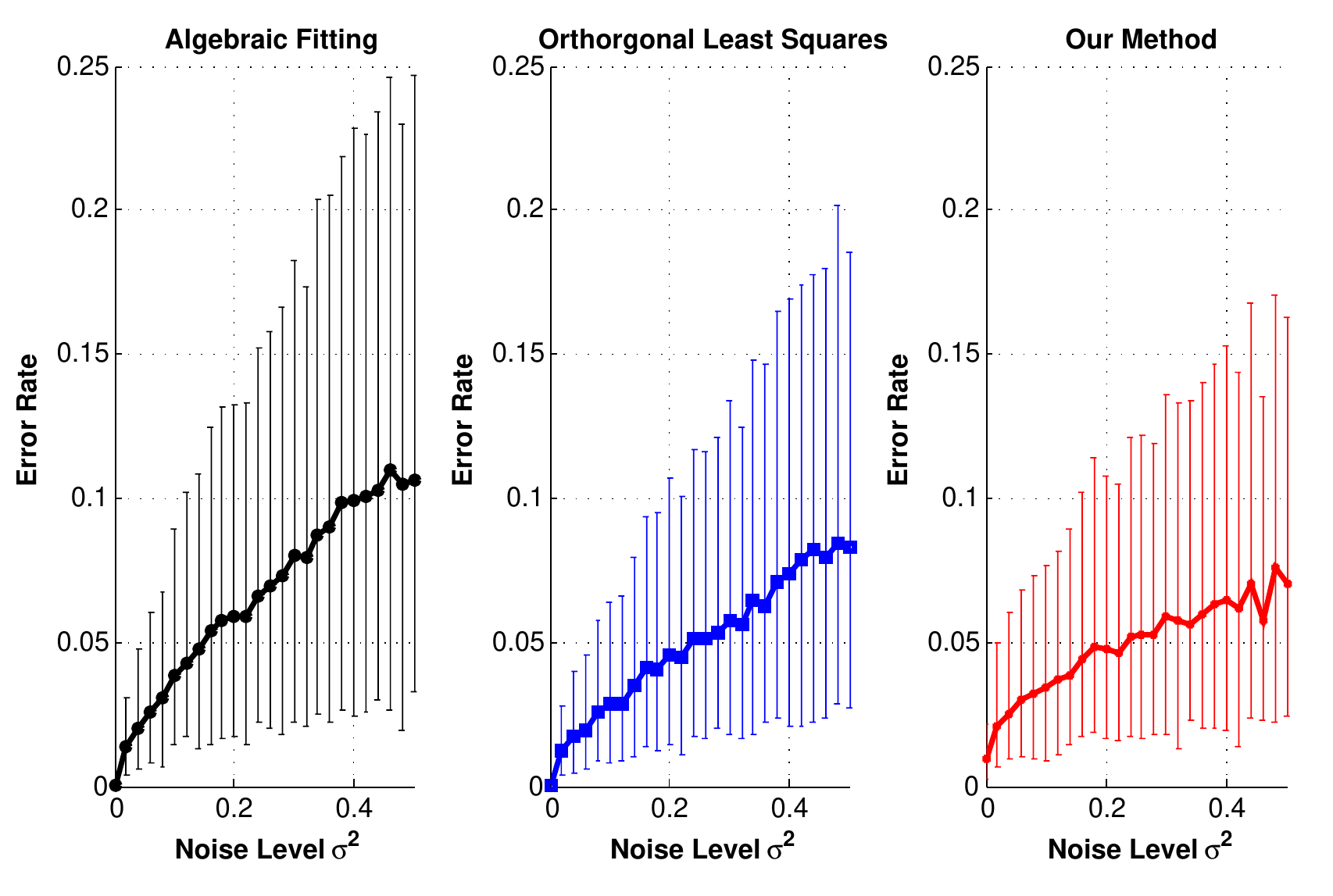}
\caption{Average error rate for algebraic fitting, OLSF, penalized fitting with 20\% quantile and 80\% quantile error bars.} \label{ConvergenceVar}
\end{figure}

Fig. \ref{ConvergenceVar} shows the comparison with error bars for the algebraic fitting, OLSF and penalized fitting. As in the previous case, the lower bounds and upper bounds of the error bars are the 20\% and 80\% quantiles of 50 trials. This demonstrates the robustness of our algorithms.

As for the computational cost, our algorithms perform almost the same as the ML algorithms and are much more efficient than the OLSF algorithms in a typical setting.

\subsection{Spheroid Fitting}
\begin{figure}[!t]
\centering
\includegraphics[width=2.3in]{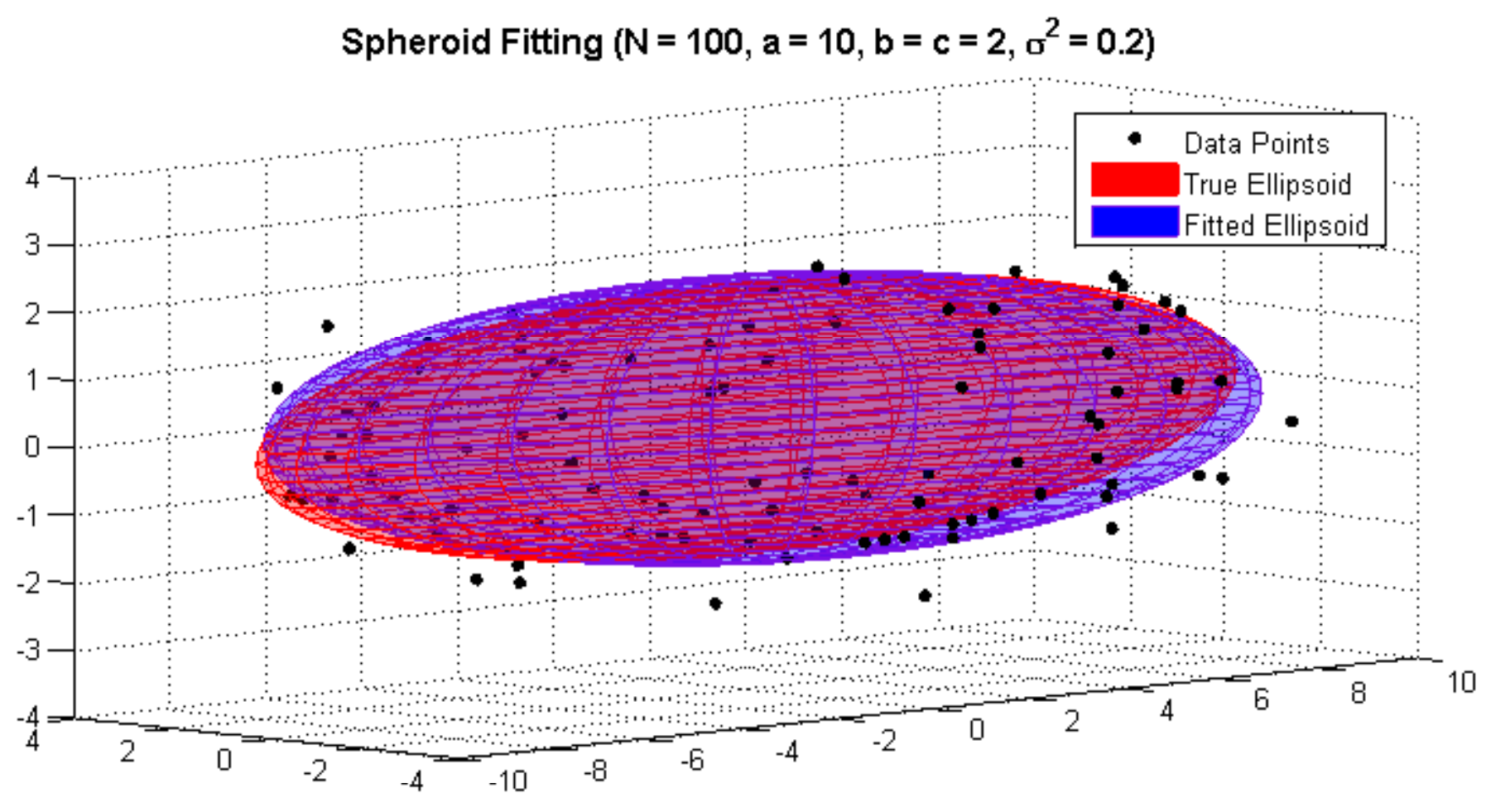}
\caption{Spheroid fitting result} \label{SpheroidSimu}
\end{figure}

Here we present an example of a typical spheroid fitting result to demonstrate the efficacy of our spheroid fitting algorithm based on the weighted objective function. Fifty true points are generated from the surface of a $10\times2\times2$ spheroid centered at the origin with rotation. The data points are generated from the true points with additive Gaussian noise with mean $\mathbf{0}$ and variance $0.2\mathbf{I}$. Fig. \ref{SpheroidSimu} shows the fitting result of our algorithm.

\bibliographystyle{IEEEtranS}
\bibliography{Asiloma}

\end{document}